# Easy-GT: Open-Source Software to Facilitate Making the Ground Truth for White Blood Cells' Nucleus

Zahra Mousavi Kouzehkanan,[a,b,⊥ *] Sajad Tavakoli,[b,c,⊥], Arezoo Alipanah[d]

[a]School of ECE, College of Engineering, University of Tehran, Tehran, Iran

[b]Nimaad Health Equipment Development Company, Tehran, Iran

[c]Faculty of Electrical Engineering, K. N. Toosi University of Technology, Tehran, Iran

[d]Faculty of Mechanical Engineering, K. N. Toosi University of Technology, Tehran, Iran

[⊥] Equally Contribution
*Corresponding author: e-mail:     z_mousavi@ut.ac.ir

**Abstract**

The nucleus of white blood cells (WBCs) plays a significant role in their detection and classification. Appropriate feature extraction of the nucleus is necessary to fit a suitable artificial intelligence model to classify WBCs. Therefore, designing a method is needed to segment the nucleus accurately. There should be a comparison between the ground truths distinguished by a hematologist and the detected nuclei to evaluate the performance of the nucleus segmentation method accurately. It is a time-consuming and tedious task for experts to establish the ground truth manually. This paper presents an intelligent open-source software called Easy-GT to create the ground truth of WBCs' nucleus faster and easier. This software first detects the nucleus by employing a new Otsu's thresholding-based method with a dice similarity coefficient (DSC) of 95.42 %; the hematologist can then create a more accurate ground truth, using the designed buttons to modify the threshold value. This software can speed up ground truth's forming process more than six times.

## 1. Introduction

There are three types of blood cells: platelets, white blood cells, and red blood cells; only white blood cells (WBCs) are responsible for providing the body's immunity. They protect the body by invading bacteria and viruses. The WBCs are first created in the bone marrow; then, they enter peripheral blood after proceeding with the maturing cycle [1]. Mature WBCs in peripheral blood are mainly categorized into five types [2]: lymphocyte, monocyte, neutrophil, eosinophil, and basophil. Fig. 1 shows these five mentioned types. Each of these types has its role in defending the body against diseases and infections. Accordingly, based on the disease the body is suffering from, the number of a specific type of WBCs either increases or decreases [2]. For example, if an individual gets vitamin deficiency, neutrophils will be decreased (neutropenia) [3]. Another good example is lymphocytosis, which means a significant increase in the number of lymphocytes in the peripheral blood that can be a symptom of leukemia [4]. So, detection, classification, and differential count of WBCs are very important for the primary diagnosis of diseases. Many researchers have addressed this subject with the help of machine learning and image processing techniques.

Artificial intelligence methods need to extract appropriate features from WBC's nucleus to detect, classify, and count WBCs. Therefore, the nucleus of the WBC must be segmented automatically using an algorithm. To evaluate the segmentation algorithm, physicians, hematologists, or experts must prepare the ground truth of nuclei manually. Therefore, identifying ground truth by an expert is an essential and critical point; otherwise, the validity of the work would be questioned. Moreover, some deep convolutional neural networks (CNN) such as U-Net [5] or SegNet [6] are used for pixel-wise segmenting the objects. Since these networks are trained supervised, the ground truths are utilized as the output of the network.

With the significant success of deep convolutional neural networks in the last decade, researchers are more inclined to use CNN-based methods than traditional image processing methods. The deep CNNs require much data to train. Designing a deep convolutional network to segment the object in the image requires many ground truths to be created. For medical and microscopic images, the ground truth must be detected by experts to be valid and reliable, which is a tedious and time-consuming task to be done manually. In various articles, there is an emphasis on making the ground truth manually by experts or physicians. Khalvati et al. [7] segmented the breast in 3D MR images. They mentioned that an expert manually diagnosed the ground truths. Dong et al. [8] have proposed a new adversarial neural network called U-net-GAN for segmenting different organs in thorax CT images. They used pre-prepared ground truths that medics detected manually. [9] and [10] used deep convolutional neural networks to extract objects in pelvis CT images. The authors of these two papers stated that experienced oncologists identified ground truths manually. Ghaneh et al. [11] have segmented WBCs in microscopic images using a combination of the K-means and the watershed algorithm. They used the ground truths generated manually by experts to evaluate the performance of the proposed algorithm. In [12] TishuQuant tool was employed to segment the WBCs' nucleus in microscopic images from peripheral blood. Hegde et al. [12] used the ground truths obtained by an expert to evaluate the proposed method. Also, in [13], with the Grabcut algorithm's help, first, WBCs were segmented from microscopic images, then the performance of the proposed method was evaluated by expert-specified ground truths.

According to the above, the importance of determining the ground truth by a physician or specialist is apparent. Due to the doctors' and hematologists' lack of time for manually preparing the ground truth, a smart tool is needed to accelerate producing it. In this paper, intelligent open-source software is presented to facilitate and speed up building the ground truth for the WBC nucleus in microscopic images. Considering that this software is open-source, modifications can make this app suitable for creating the ground truth for other medical fields.

The following section presents more details of this software and the proposed nucleus segmentation method. The following section explains more information on this software and the proposed nucleus segmentation method.

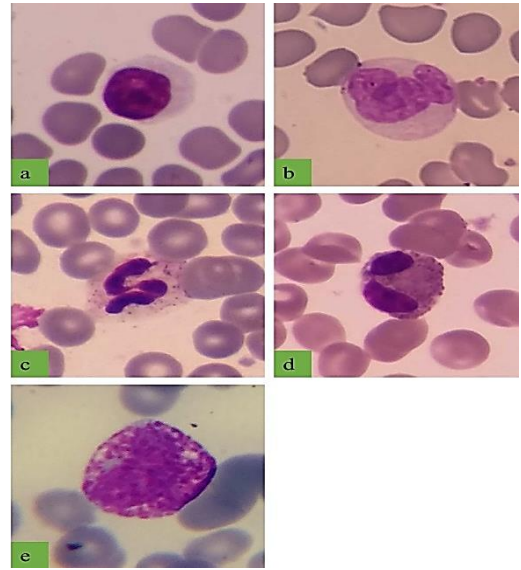

**Figure 1.** (a) Lymphocyte. (b) Monocyte. (c) Neutrophil. (d) Eosinophil. (e) Basophil

## 2. Easy-GT software

In this section, software details are explained. We have developed Easy-GT software using the python 3.7 and Tkinter library [14]. Fig. 2 shows the graphical interface of Easy-GT software. By choosing the image folder, opening it via the software, and clicking the start button, the user can quickly move between the images to edit them. At first, a primitive segmentation is applied to each image to create the nucleus ground truth; then, the final result appears to the user. The proposed segmentation is described in the following subsection, which has a threshold value as a chief hyperparameter. The default value of this threshold is chosen using the Otsu thresholding method. The user can go to the next image if the nucleus segmentation is accurate. Otherwise, only by changing the threshold (increasing it or decreasing it) can the user quickly improve the segmentation and achieve precise ground truth. As a result, the ground truth can be obtained effortlessly with just a few clicks.

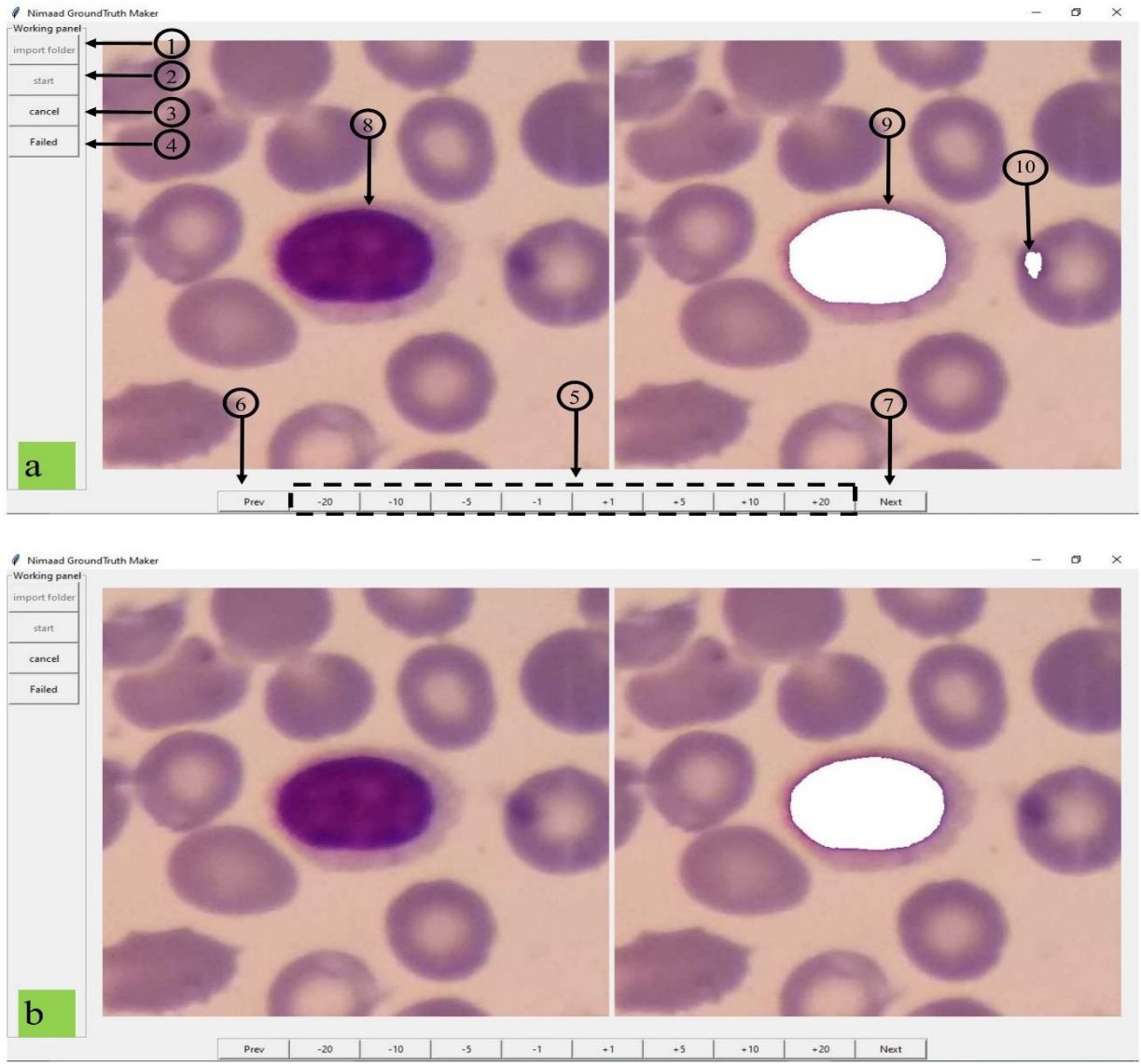

**Figure 2.** The graphical interface of Easy-GT software. In (a) the primary segmented nucleus has not been accurate; and in (b) the user has modified the ground truth by means of the buttons. (1) Selecting the folder. (2) Starting to create the ground truth. (3) Cancelling the operation of creating ground truth. (4) If ground truth is not created well, this button will save the image to a separate folder called the failed images. (5) Decreasing or increasing threshold value. (6) Next image. (7) Previous image. (8) The nucleus of a white blood cell. (9) The primitive ground truth extracted using the segmentation algorithm. (10) a part of red blood cell that wrongly detected as the ground truth.

## 3. Nucleus Segmentation Algorithm

The nucleus segmentation algorithm contains three phases, which are as follows: First, applying a color balancing algorithm to the input RGB image [15], then the color-balanced image is converted to the CMYK color space, and finally, segmenting the nucleus using a threshold value computed through otsu's thresholding algorithm (OTA) [16]. The steps of the proposed method have been shown in Fig. 3. For color balancing (first phase), the three R, G, and B channels are balanced separately. If *g* is the grayscale of the input image, the new color-balanced R, G, and B components are calculated utilizing equations (1), (2), and (3).

$$R = \frac{mean(g)}{mean(R)} \times R \quad (1)$$

$$G = \frac{mean(g)}{mean(G)} \times G \quad (2)$$

$$B = \frac{mean(g)}{mean(B)} \times B \quad (3)$$

After examining different color spaces in the second phase, we concluded that the M channel of CMYK color space is more discriminative for the WBC's nucleus (Fig. 3). The CMYK color space includes four channels called cyan (C), magenta (M), yellow (B), and black (K). In below, Equations (4) to (10) shows the steps of converting RGB color space to CMYK color space [17]:

$$C' = 255 - R \quad (4)$$

$$M' = 255 - G \quad (5)$$

$$Y' = 255 - B \quad (6)$$

$$K = \min(C', M', Y') \quad (7)$$

$$C = \frac{C' - K}{255 - K} \quad (8)$$

$$M = \frac{M' - K}{255 - K} \quad (9)$$

$$Y = \frac{Y' - K}{255 - K} \quad (10)$$

In the third phase, Otsu's thresholding algorithm (OTA) was used. OTA was presented in 1979 by Noboyuki Otsu [16]. This algorithm finds a suitable threshold value by considering the statistical distribution of the image histogram. The Otsu's algorithm has successfully segmented medical images because the histogram of these images usually consists of two meaningful peaks representing two different ranges of pixels. This algorithm is adopted to regulate an adaptive threshold for each image. The optimal threshold is between 0 and 255. So, the highest between-class variance is for a set of pixels above this threshold or below it. This adaptive threshold helps the user to spend less time to get the actual ground truth. For this paper, two-class-OTA and three-class-OTA were employed. The two-class-OTA splits the image histogram into two categories and calculates one threshold value, named *THV1*, to distinguish between them. The three-class-OTA divides the image histogram into three classes with low, medium, and high pixel intensity. The borders of which are specified by two threshold values (upper and lower threshold values). The upper threshold value was named *THV2*. Afterward, a convex combination of *THV1* and *THV2* was considered to calculate the ultimate threshold value (*UTHV*). This approach was formulated by α parameter (equation (11)). The optimal value for α was obtained by varying its amount between 0 and 1 with a step of 0.1 and calculating the dice similarity coefficient of 250 WBCs' images for each α value. The best α value was equal to 0.3. Fig. 4 illustrates *THV1*, *THV2*, and *UTHV*.

$$UTHV = \alpha \times THV1 + (1 - \alpha) \times THV2, \ 0 \leq \alpha \leq 1 \quad (11)$$

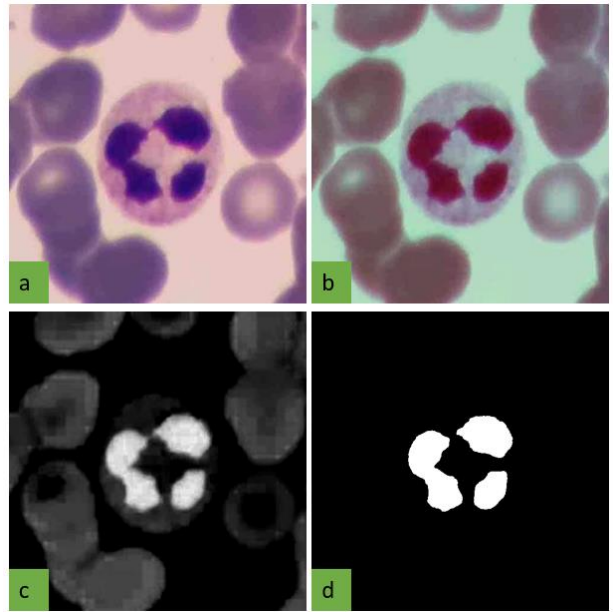

**Figure 3**. The steps of the proposed method. (a) Original RGB image. (b) Color balanced RGB image. (c) M channel of CMYK color space. (d) Segmented nucleus

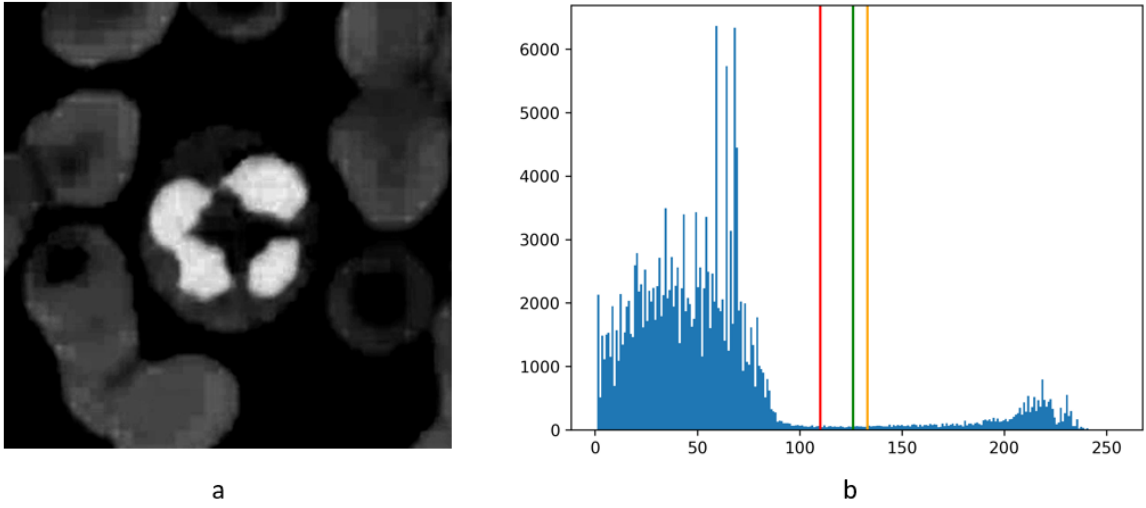

**Figure 4.** (a) M channel. (b) M channel's histogram (pixels with zero intensity were ignored); the red line is *THV1*, the orange line is *THV2*, and the green line is *UTHV* obtained from equation (11).

## 4. Results

### 4.1. Performance of the Nucleus Segmentation Algorithm

The ground truths of 250 images, including 50 images from five classes (lymphocyte, monocyte, neutrophil, eosinophil, and basophil), were identified by an expert to evaluate the efficiency of the proposed nucleus segmentation method. These images were collected by Nimaad Health Development Company, settled in Tehran, Iran. The Olympus CX18 microscope and camera phone of the Samsung Galaxy S5 were used to gather these images. Three criteria of sensitivity, precision, and dice similarity coefficient (DSC) were calculated for each of α values to assay α variation's effect. Forasmuch as the DSC criterion is better than sensitivity and precision, only DSC was considered to find the optimal value for α.

Fig.4 shows that the perfect value for α based on DSC is equal to 0.3. Therefore, considering the 0.3 value for α, the proposed nucleus segmentation method can detect the nuclei with sensitivity, precision, and DSC of 98.27 %, 93.62 %, and 95.42 %, respectively. The definitions of the mentioned criteria have been presented in relations (12) to (14). TP, FP, FN are abbreviations for true positive, false positive, and false negative, consecutively.

By taking a detailed look at Figure 5, it is seen that as the alpha value gains, the sensitivity increases, and the precision decreases. With the enhancement in alpha value, the *UTHV* becomes closer to *THV1* and drops, and more pixels become segmented. Thus, according to equations 12 and 13, the sensitivity and precision boost and fall, respectively.

$$Sens = \frac{TP}{TP+FN} \quad (12)$$

$$Prec = \frac{TP}{TP+FP} \quad (13)$$

$$DSC = 2 \times \frac{TP}{(TP+FP)+(TP+FN)} \quad (14)$$

### 4.2. Accelerating the Ground Truth Making Process

Working with this software shows that it can speed up the ground truth production process significantly. An expert was asked to create the ground truths of WBCs' nucleus with and without this software for one hour. The expert could manually generate only 55 ground truths (utilizing a tablet and a pen), whereas he made 337 ground truths using this software during the same amount of time. By dividing 337 to 55, it can be seen that this software can accelerate the process of making ground truths more than six times.

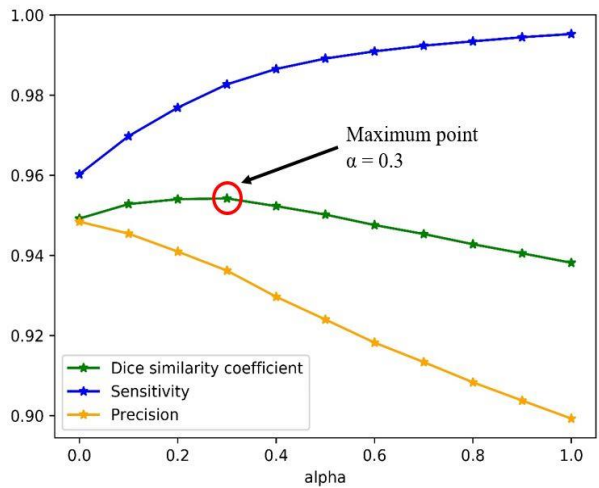

**Figure 5.** Effect of α value variation on selected criteria

**Table 1.** Sensitivity, precision, and dice similarity coefficient percentage for different $\alpha$ values

| Criteria  α | Sens (%) | Prec (%) | DSC (%) |
|---|---|---|---|
| α = 0 | 96.02 | 94.84 | 94.19 |
| α = 0.1 | 96.98 | 94.54 | 95.28 |
| α = 0.2 | 97.69 | 94.10 | 95.40 |
| **α = 0.3** | **98.27** | **93.62** | **95.42** |
| α = 0.4 | 98.65 | 92.97 | 95.23 |
| α = 0.5 | 98.91 | 94.00 | 95.02 |
| α = 0.6 | 99.09 | 91.82 | 94.76 |
| α = 0.7 | 99.23 | 91.34 | 94.53 |
| α = 0.8 | 99.34 | 90.83 | 94.28 |
| α = 0.9 | 99.45 | 90.38 | 94.05 |
| α = 1 | 99.53 | 89.93 | 93.82 |

## 5. A Comparison with the State-of-the-Arts

This section compares the Easy-GT software segmentation approach with three other methods used in our previous works [2]: U-Net++ [18], attention U-Net [19], and Tavakoli *et al.'s* method [2]. U-Net++ and attention U-Net are two deep convolutional neural networks derived from the U-Net model [5]. Since training these two models is a supervised process, several adequate ground truths must be provided for learning them.

Therefore, almost 1000 images from the Raabin-WBC dataset [20] were randomly selected, proceeded by identifying their ground truths. These images were employed to train the models, and the 250 images used to evaluate this paper's method were also applied for U-Net++ and attention U-Net's evaluation. These two models were learned with 40 epochs. Table 2 illustrates the comparisons [2].

This table shows that the proposed method's performance is near the U-Net++ and attention U-Net, whereas it is significantly faster than the aforesaid deep model. The introduced method can averagely segment the nucleus within only 47 milliseconds, while this operation takes 1612 and 628 milliseconds for U-Net++ and attention, respectively [2].

Besides, our method has no trainable parameters, but U-Net++ and attention U-Net has more than half a million trainable parameters. All the methods (this paper's method, previous work, U-Net++, and attention U-Net) were executed on Google Colab.

**Table 2.** Comparison of the proposed method with other works.

| Method | DSC (%) | Time (ms) | Trainable parameters |
|---|---|---|---|
| Tavakoli *et al.* [2] | 96.75 | 45 | 0 |
| U-Net++ [18] | 97.19 | 1612 | 897412 |
| Attention U-Net [19] | 96.33 | 628 | 854936 |
| Proposed method | 95.42 | 47 | 0 |

## 6. Code Availability

The codes of this software are available at the below address:
https://github.com/nimaadmed/Easy-GT

## 7. Discussion and Conclusion

In this research, to accelerate and facilitate preparing the ground truth of WBCs' nucleus by hematologists, a new software was presented, which is very user-friendly and easy to work. After making the ground truth of the nucleus by the proposed segmentation algorithm, the user can easily modify it by clicking the designed buttons and changing the primary threshold value obtained from the segmentation algorithm. As mentioned before, this software can accelerate the ground truth forming process more than six times. Nimaad Health Equipment Development Company (Nimaadmed.com) currently uses this software and has provided the required data. Since this software is open-source, researchers and developers worldwide can modify the software and improve it to suit their needs.

## Author contributions

## Competing interests